\documentclass[letterpaper]{article} 
\usepackage{aaai25}  
\usepackage{times}  
\usepackage{helvet}  
\usepackage{courier}  
\usepackage[hyphens]{url}  
\usepackage{graphicx} 
\usepackage{natbib}  
\usepackage{caption} 
\usepackage{algorithm}
\usepackage{algorithmic}
\usepackage{subcaption}
\usepackage{amsmath}

\DeclareMathOperator*{\argmin}{arg\,min}
\usepackage{amsfonts}
\usepackage{newfloat}
\usepackage{listings}
\DeclareCaptionStyle{ruled}{labelfont=normalfont,labelsep=colon,strut=off} 
\floatstyle{ruled}
\newfloat{listing}{tb}{lst}{}
\floatname{listing}{Listing}
\newcommand{\system}{{MalCL}}

\title{MalCL: Leveraging GAN-Based Generative Replay to Combat Catastrophic Forgetting in Malware Classification}

\author{
    Jimin Park\textsuperscript{\rm1},
    AHyun Ji\textsuperscript{\rm1},
    Minji Park\textsuperscript{\rm1},
    Mohammad Saidur Rahman\textsuperscript{\rm2},
    Se Eun Oh\textsuperscript{\rm1}\thanks{corresponding author}}
    
\affiliations{
    \textsuperscript{\rm 1}Ewha Womans University, 
    \textsuperscript{\rm 2}University of Texas at El Paso\\

    242aig12@ewha.ac.kr, nolli77@ewha.ac.kr, 02mjpark@ewha.ac.kr, seoh@ewha.ac.kr, msrahman3@utep.edu
}

\usepackage{bibentry}

\usepackage{todonotes}
\usepackage{xcolor}
\usepackage{colortbl}
\usepackage{multirow}
\usepackage{booktabs}

\begin{document}

\maketitle



\begin{abstract}




Continual Learning (CL) for malware classification tackles the rapidly evolving nature of malware threats and the frequent emergence of new types. Generative Replay (GR)-based CL systems utilize a generative model to produce synthetic versions of past data, which are then combined with new data to retrain the primary model. Traditional machine learning techniques in this domain often struggle with catastrophic forgetting, where a model's performance on old data degrades over time.

In this paper, we introduce a GR-based CL system that employs Generative Adversarial Networks (GANs) with feature matching loss to generate high-quality malware samples. Additionally, we implement innovative selection schemes for replay samples based on the model’s hidden representations.

Our comprehensive evaluation across Windows and Android malware datasets in a class-incremental learning scenario -- where new classes are introduced continuously over multiple tasks -- demonstrates substantial performance improvements over previous methods. For example, our system achieves an average accuracy of 55\% on Windows malware samples, significantly outperforming other GR-based models by 28\%. This study provides practical insights for advancing GR-based malware classification systems. The implementation is available at \url {https://github.com/MalwareReplayGAN/MalCL}\footnote{The code will be made public upon the presentation of the paper}.

\end{abstract}

\section{Introduction}
\label{sec:intro}
Machine learning (ML) has become essential for protecting computers and networked systems, with successful applications in malware detection and classification across domains like Windows and Android~\cite{arp2014drebin,dahl2013large,raff2021classifying, malwareguard, maiorca2012pattern}. Researchers have developed advanced models that distinguish malware from benign software, classify malware families, and categorize malware types by analyzing features from both benign and malicious source codes.

Malware evolves rapidly, requiring frequent retraining on large datasets to stay effective. Daily, the AV-TEST Institute logs 450,000 new malware and ``Potentially Unwanted Applications (PUA)''~\cite{av-test}, and VirusTotal processes over one million software submissions~\cite{virustotal}, posing significant challenges for antivirus companies. In response, vendors face tough choices: remove old samples from training sets, risking the resurgence of older malware; overlook new samples, missing emerging malware trends; reduce training frequency, sacrificing accuracy; or invest heavily in continuous retraining.

To address these challenges, the approach suggested by Raff et al.~\cite{raff2021classifying} provides a viable method to enhance memory efficiency in training malware classification models. However, this method does not tackle the inherent tendency of ML models to forget previously learned malware feature distributions when they are continuously trained with new data -- a necessity driven by the dynamic and incremental nature of the malware domain.

Continual Learning (CL) offers an effective solution to this challenge, commonly referred to as Catastrophic Forgetting (CF). CL continually adapts to a data stream, refining knowledge over time through techniques such as \textit{replay} (i.e., storing and reusing data) and \textit{regularization} (i.e., regularizing learned representation). These strategies help reduce the storage and computational burdens associated with frequent retraining~\cite{continual-learning-malware,channappayya2024augmented}.


In particular, among replay techniques~\cite{cao2024generative, bir, icarl}, Generative Replay (GR)~\cite{gr}, which utilizes generative deep neural networks to produce synthetic samples that represent past data distributions, is advantageous since GR approach eliminates the need to store past raw data, offering a significant benefit in scenarios where data storage is constrained due to privacy regulations.

While CL techniques are well-established in computer vision (CV)~\cite{hsu2018re,bir}, their application in malware classification remains underdeveloped. \citet{continual-learning-malware} demonstrated that CL methods tailored for CV often perform poorly in malware detection due to the complex and diverse nature of malware features~\cite{continual-learning-malware}. They found that directly applying CV-specific CL systems to malware classification without considering the complexities of malware data distribution leads to poor detection of known malware classes, performing no better than the ``{\em None}'' baseline, which involves retraining the model with only new data.

In this paper, we introduce \system, a novel malware family classification system that utilizes a GR-based CL approach, incorporating a Generative Adversarial Network (GAN) architecture for its replay methodology~\cite{goodfellow2020generative} and a one-dimensional Convolutional Neural Network (CNN) for classifying malware families. We have implemented feature matching in the design of the loss function for the GAN's generator to enhance the quality of generated samples. Additionally, we developed several replay sample selection techniques to select the most effective samples for each malware class. 

MalCL is designed to accurately classify malware families using feature vectors in a class-incremental learning scenario, as outlined in prior research~\cite{van2022three}, addressing multiple continual tasks with each introducing new malware classes (or families) and effectively classifying both past and newly encountered malware families. The primary goal of this system is to mitigate CF of previously observed malware classes while efficiently adapting to new malware families.

We summarize our key contributions as follows:

\begin{itemize}

    \item \textbf{Malware Domain-Specific CL Model:} We introduce MalCL, a novel CL model tailored specifically for malware family classification. It achieves an average accuracy of 55\% in correctly identifying 100 malware families across 11 CL tasks, each introducing new malware families. This performance surpasses existing approaches such as Generative Replay (GR)~\cite{gr} and  Brain-Inspired Replay (BI-R)~\cite{bir} by 28\%. Our model incorporates robust architectures designed for precise malware classification.
    
    \item \textbf{Improved Replay with Feature Matching:} MalCL effectively replays past malware samples using a GAN’s generator. We have enhanced this capability by incorporating Feature Matching Loss (FML), which reduces the discrepancy between the richer features of original malware samples -- extracted from the hidden layers of the GAN’s discriminator -- and the synthetic malware samples created by the GAN's generator.
    
    \item \textbf{Replay Sample Selection Schemes:} We have developed innovative replay sample selection schemes for MalCL, particularly focusing on features derived from the intermediate layer of the classifier. This approach improves the alignment between generated malware samples and original data, thereby improving classification accuracy.
    
    \item \textbf{Strategic Task Set Construction:} We have explored various task set construction strategies, especially assigning larger classes to initial tasks, which has optimized MalCL’s performance and showcased a strategic approach to effectively mitigate CF in malware classification domain.
    
\end{itemize}

\section{Threat Model}

\begin{figure}[!t]
    \centering
    \includegraphics[width=0.8\linewidth]{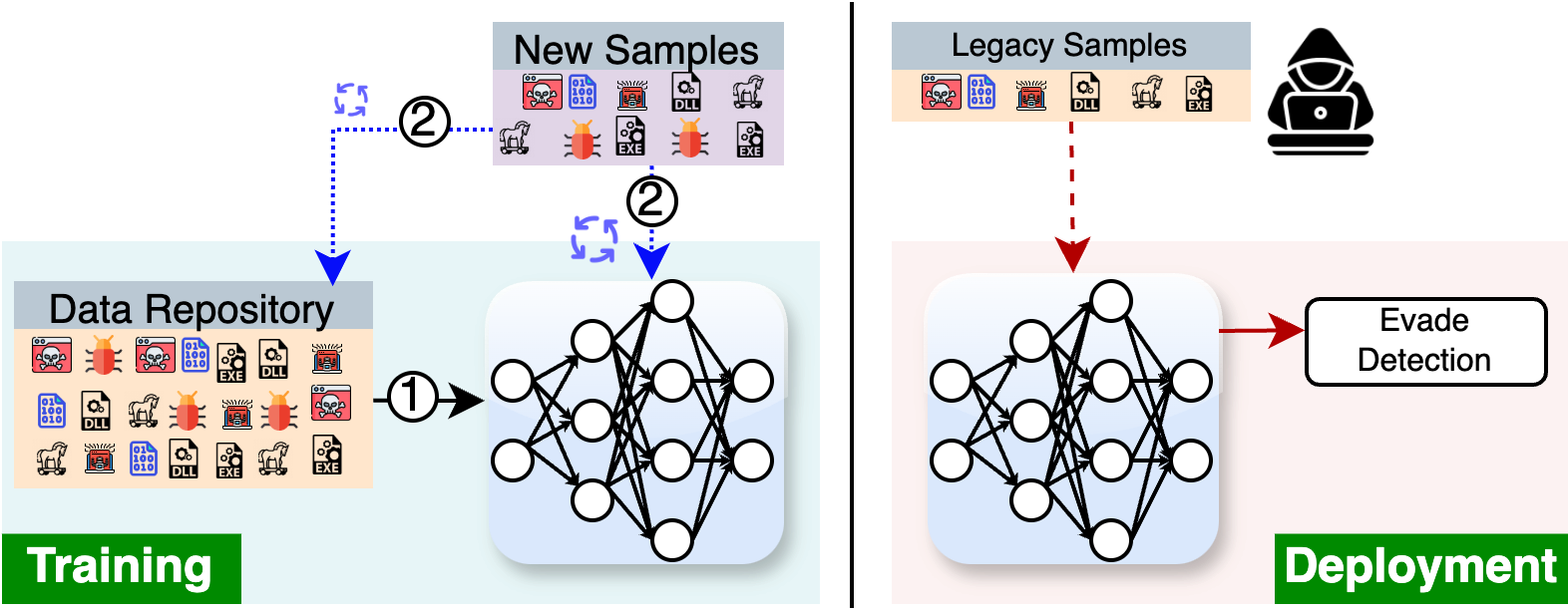}
    \caption{An attacker reuses legacy malware to evade systems updated with {\em only} new malwares.}
    \label{fig:CFthreatModel}
    \vspace{-0.4cm}
\end{figure}

A schematic representation of the threat model, where an attacker reuses legacy malwares to exploit vulnerabilities in machine learning (ML) systems, is shown in Figure~\ref{fig:CFthreatModel}. {\Large \textcircled{\normalsize 1}} in Figure~\ref{fig:CFthreatModel} represents the initial phase, where the system is trained on available data from the repository and deployed. When new samples are observed, the system is retrained with these samples, as depicted by {\Large \textcircled{\normalsize 2}}, and the new samples are added to the data repository. However, due to catastrophic forgetting (CF), the system's ability to detect earlier threats diminishes over time.

This limitation allows attackers to reuse older malware and malicious code snippets, which can bypass the updated system. The likelihood of successful evasion increases as the temporal gap between the original training and the attack widens. Therefore, addressing this challenge is critical to developing defense systems that maintain consistent detection capabilities for both recent and legacy malware.
\section{Related Work}
\label{sec:related}

The fundamental challenge in developing a continual learning (CL) system is addressing {\em catastrophic forgetting} (CF). CF is a phenomena in which the performance of a neural network degrades significantly on older tasks after being trained on a new task~\cite{french1999catastrophic, mccloskey1989catastrophic}. Over the years, several mechanisms have been proposed to overcome CF~\cite{si,bir,icarl}, which fall into two major categories~\cite{parisi2019continual}: replay methods and regularization methods. Replay techniques supplement the training data for each new task with representative data from previously learned tasks.

\paragraph{Replay.} Replay techniques can be further classified into major two subcategories: exact replay and generative replay. Exact replay methods, such as Experience Replay (ER)~\cite{er} and Incremental Classifier and Representation Learning (iCaRL)~\cite{icarl}, allocate a memory budget $\mathcal{M}$ to replay samples from previous tasks, combining them with new data during retraining to optimize performance. Generative replay approaches, such as Generative Replay (GR)~\cite{gr}, Brain-Inspired Replay (BI-R)~\cite{bir}, diffusion-based generative replay (DDGR)~\cite{gao2023ddgr} generate representation of the stored samples with a secondary generative models to generate previous data distributions, making them useful when access to historical data is limited.


\paragraph{Regularization.} Regularization-based methods, such as Elastic Weight Consolidation (EWC)\cite{ewc} and Synaptic Intelligence (SI)\cite{si}, work by limiting changes to weights that are crucial for previously learned tasks. These methods introduce an additional loss function, known as regularization loss, which is added to the training loss. The total loss therefore reflects both the new learning and the retention of knowledge from previous tasks. This approach helps strike a balance between acquiring new information and preserving important past learnings, thereby enhancing the system's adaptability.




\subsection{CL and Related Approaches for Malware Classification}

Few studies have explored the application of CL in malware detection. \citet{continual-learning-malware} are among the first to investigate CL for malware classification, finding that existing methods fall short in effectively addressing CF. More recently, \citet{chen2023continuous} combined contrastive learning with active learning to train Android malware classifiers, focusing on detecting concept drift rather than overcoming CF. \citet{channappayya2024augmented} explored a replay-based CL technique in network intrusion detection, which involves class-balancing reservoir sampling and perturbation assistance, though this domain differs from the challenges faced in malware classification.

Other approaches, such as {\em online learning}, have been used for malware classification to incorporate new samples as they appear, but they do not directly tackle CF~\cite{droidevolver}. Transfer learning has also been studied for malware detection due to its adaptability to evolving threats~\cite{neyshabur2020being}, but it often neglects the retention of knowledge from previous tasks, which is crucial to avoid the resurfacing of vulnerabilities~\cite{chai2022dynamic}. Considering these limitations, our paper focuses on CL, which is specifically designed to address CF in malware detection.

\begin{figure}[!t]
    \centering
    \includegraphics[width=0.4750\textwidth]{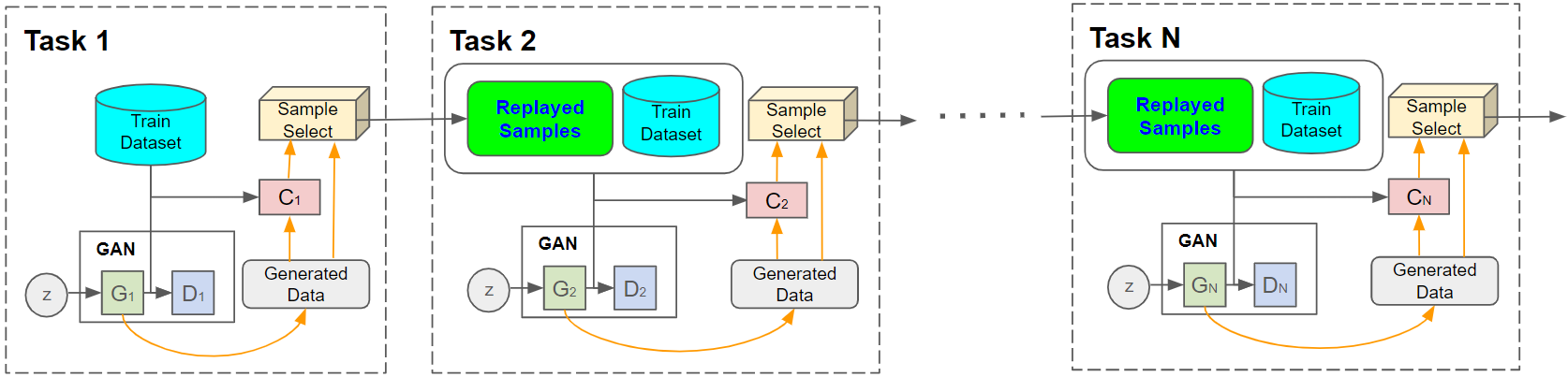}
    \caption{MalCL continual learning pipeline for malware classification using N tasks.}
    \label{fig:pipeline}
    \vspace{-0.3cm}
\end{figure}

\section{MalCL Overview}
\label{sec:system}

In this section, we detail our proposed \system, which aims to accurately detect malware families based on features extracted from Windows Portable Executable (PE) and Android APK files.

\subsection{MalCL Pipeline}

A schematic representation of the MalCL pipeline is depicted in Figure~\ref{fig:pipeline}. MalCL which consists of two primary components: a GAN and a classifier. The GAN is responsible for generating malware samples of the training data from previous tasks to be replayed, while the classifier is tasked with malware family classification, identifying the family to which each feature vector belongs. It is important to note that the feature vectors utilized in this study were constructed based on previous literature~\cite{continual-learning-malware}.


As depicted in Figure~\ref{fig:pipeline}, throughout each of the $n$ tasks, the GAN is trained with a generator ($G_i$) designed to produce high-quality synthetic malware samples that closely resemble the training data of the $i$-th task. Concurrently, a discriminator ($D_i$) works to differentiate these generated samples from the actual training data. The classifier model ($C_i$) is trained using the training data to accurately identify malware families. Once $G_i$ produces sufficiently realistic malware samples, these are fed into $C_i$, which then outputs \texttt{logits} from the middle layer of the classifier model.


Subsequently, a replayed sample set is created by selecting the most effective synthetic malware samples using a selection algorithm. We refer to these samples as ``replay samples'' throughout the paper. Once the replayed set is established, it is carried forward to the next task, where it is used alongside the new training data to train both the GAN and the classifier models.



\subsection{Overview of the Models} 
In this section, we detail the architectures of the GAN and classifier models chosen for our study.

\subsubsection{GANs.} 

As shown in Figure~\ref{subfig:generator}, the generator comprises four 1D CNN layers, two Fully-Connected (FC) layers, and three deconvolution layers, with ReLU activation and batch normalization applied to all but the final deconvolution layer. The discriminator (Figure~\ref{subfig:disc}) includes two convolutional layers followed by two FC layers, with ReLU activation and batch normalization in all but the final FC layer. Flattened logits from the second convolutional layer are used for feature matching to analyze differences between fake and original malware samples. A sigmoid layer serves as the output for both the generator and discriminator. Optimization is performed using Adam with a batch size of 256.


\subsubsection{Classifier.}

\begin{figure}[!t]
  \centering
    \begin{subfigure}[b]{0.30\linewidth}
        \includegraphics[width=\linewidth]{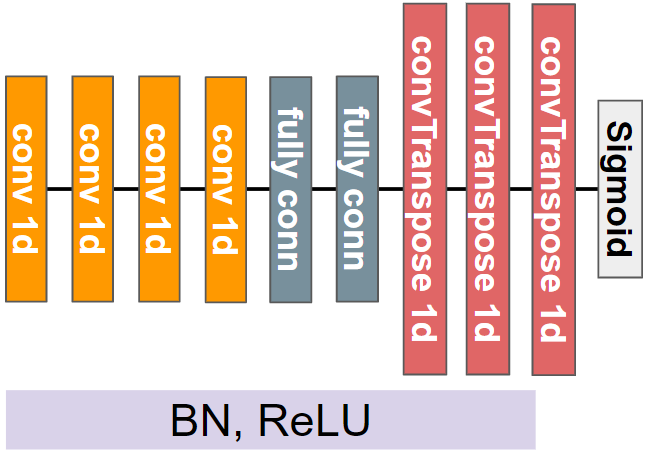}
        \caption{Generator}
        \label{subfig:generator}
    \end{subfigure}
    \hspace{0.05\linewidth}
    \begin{subfigure}[b]{0.2\linewidth}
        \includegraphics[width=\linewidth]{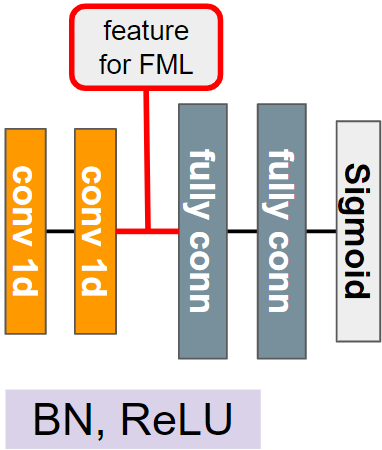}
        \caption{Discriminator}
        \label{subfig:disc}
    \end{subfigure}
    \hspace{0.05\linewidth}
    \begin{subfigure}[b]{0.25\linewidth}
        \includegraphics[width=\linewidth]{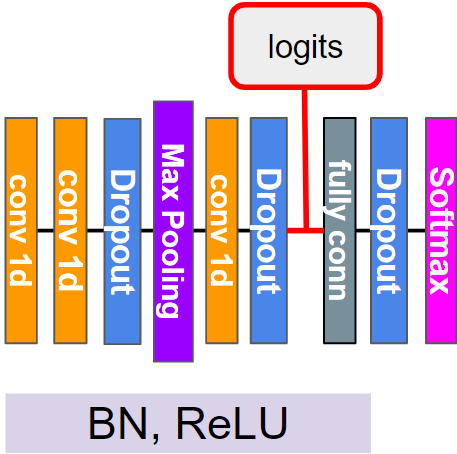}
        \caption{Classifier}
        \label{fig:classifier}
    \end{subfigure}

    \caption{GAN architecture: (a) Generator and (b) Discriminator, adapted for generating replay malware samples, and (c) Classifier, optimized for malware family classification in continual learning tasks.}
    \label{fig:gan}
    \vspace{-0.3cm}
\end{figure}

The classifier model in Figure~\ref{fig:classifier} consists of three convolutional layers followed by a FC layer. Max pooling and dropout follow the first two convolutional layers, while dropout is applied after the third convolutional layer and the FC layer. The final output is a softmax layer. Flattened logits from the third convolutional layer are used for relay sample selection. The model is optimized using Stochastic Gradient Descent (SGD) with a learning rate of \( e^{-3} \), momentum of 0.9, and weight decay of \( e^{-7} \), with a batch size of 256.

\subsection{Replay Sample Selection Schemes}
\label{subsec:sel_algo}

To maximize the effectiveness of continual learning in MalCL, selecting high-quality synthetic malware samples is crucial. A robust selection metric must pinpoint samples that closely resemble those in the original malware training set. Additionally, the generator's tendency to produce imbalanced samples per class can lead to model collapse, with only a few classes represented by synthetic samples. To address these challenges, we propose several replay sample selection methods in this section.

Let \( F(s) \) represent a transformation applied to a synthetic sample \( s \), where \( s \in S \) and \( S \subseteq \mathbb{R}^l \), with \( l \) denoting the dimension of the generator's output layer. \( H(x) \) represents a similar transformation applied to an original sample \( (x, y) \), where \( x \in X \), \( y \in Y \), \( X \subseteq \mathbb{R}^m \), and \( Y \subseteq \{0,1\}^n \), representing the training set \( M \). The feature vector length is \( m \), and \( n \) is the total number of malware classes. The distance \( \delta \) between these transformations is calculated as:

\begin{equation}
\delta(F(s), H(x))
\end{equation}



To implement the \(k\)-Nearest Neighbor (\(k\)-NN) approach, we select the \(k\) synthetic samples in \(S\) with the lowest distances to the original samples \(x\) from each class \(c\), represented as:

\begin{equation}
    S_{c,k} = \{s  \mid  \argmin_{s} \delta(F(s), H(x)), \forall s \in S, \forall x \in X_c\}
\end{equation}


where \(X_c\) includes the original samples from malware class \(c\), and \( \delta \) denotes a distance metric, such as L1 or L2.



In addition to the \(k\)-NN approach, we also propose a global selection across \(n\) classes, aimed at selecting samples that \textit{generally} resemble the overall malware data distribution. This approach is formulated as:

\begin{equation}
    S_{km} = \{s \mid \argmin_{s} \delta(F(s), H(x)), \forall s \in S, \forall x \in X\}
\end{equation}

where \(m\) is the number of batches.


Note that \(S_{km}\) does not explicitly enforce a balanced selection per class as \(S_{c,k}\) does, which may result in zero samples for certain classes. Thus, this methodology has the limitation that the selection scheme may not fully overcome the mode collapse problem of the generator. However, we explore this scheme to demonstrate the impact of the generator's ability to tune toward producing samples that look good overall, rather than targeting each class specifically, even though it produces synthetic samples for only a subset of \(n\) classes.


These strategies construct an effective replay set and ensures class balance, addressing the generator's tendency to produce imbalanced samples. This section outlines selection algorithm adjustments to enhance replay sample effectiveness, compensating for training data loss in previous tasks. We modify transformations \( F \), \( H \), and metric \( \delta \) to achieve this.



\subsubsection{Selection with L2 Distance to One-Hot Labels.} In this setup, we use L2 distance (\(\delta\)) with one-hot-encoded labels. When a synthetic sample \(s\) is input into the classifier for the \(i\)-th task, the outputs from the final softmax layer are:
\begin{equation}
    F(s) = C_i(s), \quad H(x) = y
\end{equation}
where \(C_i(s)\) is the classifier's softmax output, and \(y\) is the true label vector.

The selection process computes the Euclidean distance between \(C_i(s)\) and \(y\), selecting the top \(k\) synthetic samples for each class \(c\) based on:
\begin{equation}
    S_{c,k} = \{s \mid \argmin_{s} \sqrt{(C_i(s)-y_c)^2}, \forall s \in S, y_c \in Y\}
\end{equation}
where \(y_c\) denotes the label vector for class \(c\).
These selected samples are then used to train the next classifier model, \(C_{i+1}\).




\subsubsection{Selection with L1 Distance to Logits.} In this setup, logits \(\mathcal{L}_{i}(x)\) and \(\mathcal{L}_{i}(s)\) are extracted from the middle layer of the \(i\)-th classifier \(C_i\) when \(x\) (from the \(i\)-th task training set \(X_i\)) and \(s\) are input. The selection process computes the L1 distance (\(\delta\)) between these logits and selects either the top \(k\) samples with the shortest distances for each class \(c\) or the top \(km\) samples across \(n\) classes using \(m\) batches. These samples are then used to train the subsequent classifier, \(C_{i+1}\).

We explored two logit representations for \(H\):


\begin{enumerate}
\item \textbf{Per-Batch Mean Logit Vector}: The mean logit vector \(\bar{\mathcal{L}}_{i,b}(x)\) for batch \(b\) with size \(n_b\) is given by:
\begin{equation}
    \bar{\mathcal{L}}_{i,b}(x) = \frac{1}{n_b} \sum_{x \in X_b} \mathcal{L}_{i}(x)
\end{equation}
Then, \(F(s)\) and \(H(x)\) are defined as:
\begin{equation}
    F(s) = \mathcal{L}_i(s), \quad H(x) = \bar{\mathcal{L}}_{i,b}(x)
\end{equation}
The top \(km\) samples across classes are selected by minimizing:
\begin{equation}
    S_{km} = \{s \mid \argmin_{s} |\mathcal{L}_i(s) - \bar{\mathcal{L}}_{i,b}(x_j)|, \forall s \in S, \forall x_j \in X_b\}
\end{equation}
where \(m\) is the number of batches, and \(X_b\) is the batch, with \(X_b \in X_i\). Finally, each \(s\) is labeled with the class label outputted by \(C_i(s)\). This scheme selects the synthetic samples closer to the global mean, that is the centroids of each batch, and adopts the natural decision of the classifier for labeling.

    \item \textbf{Per-Class Mean Logit Vector}: The mean logit vector \(\bar{\mathcal{L}}_{i,c}(x)\) for class \(c\) with size \(n_c\) is:
\begin{equation}
    \bar{\mathcal{L}}_{i,c}(x) = \frac{1}{n_c} \sum_{x \in X_c} \mathcal{L}_{i}(x)
\end{equation}
Then, \(F(s)\) and \(H(x)\) are:
\begin{equation}
    F(s) = \mathcal{L}_i(s), \quad H(x) = \bar{\mathcal{L}}_{i,c}(x)
\end{equation}
The top \(k\) samples for each class \(c\) are selected by minimizing:
\begin{equation}
    S_{c,k} = \{s \mid \argmin_{s} |\mathcal{L}_i(s) - \bar{\mathcal{L}}_{i,c}(x_j)|, \forall s \in S, \forall x \in X_c\}
\end{equation}

Compared to the former, this approach selects replay samples closer to the centroid of each class and labels them according to the closest centroid. With this approach, we achieve \(k\) samples for each class, leading to a balanced dataset.
\end{enumerate}


These selection methodologies are aimed at effectively training the classifiers by refining the selection of synthetic samples through sophisticated distance calculations.

\subsection{Loss Functions For Generator}
\label{subsec:loss}
To ensure the quality of replay malware samples, selecting an appropriate loss function for the generator is essential. In this section, we discuss two loss functions for the generator. We detail the training process of GAN in Algorithm~\ref{alg:gan}.

\begin{algorithm}[!t]
\caption{Training a GAN in MalCL}
\label{alg:gan}
\begin{algorithmic}
\STATE {Initialize \( G \) and \( D \) networks with random weights}
\STATE {Define batch size \( n \), learning rate \( \eta \), and number of epochs \( E \)}
\FOR{\( epoch = 1 \) to \( E \)}
    \FOR{each batch \( B \) in the training dataset}
        \STATE {Sample \( m \) noise samples \( \{z_1, ..., z_m\} \) from noise prior \( p_z(z) \).}
        \STATE {Generate synthetic malware samples \( \{x_1, ..., x_m\} \) where \( x_i = G(z_i) \).}
        \STATE {Sample \( m \) real malware instances \( \{x'_1, ..., x'_m\} \) from malware data distribution \( p_{data}(x) \).}
        \STATE {Update \( D \) by ascending its stochastic gradient:}
        \[
        \nabla_{\theta_d} \frac{1}{m} \sum_{i=1}^{m} \left[\log D(x'_i) + \log (1 - D(G(z_i)))\right]
        \]
        \STATE {Sample \( \{z_1, ..., z_m\} \) from \( p_z(z) \).}
        \STATE {Update \( G \) by descending its stochastic gradient:}
        \[
        \nabla_\mathbf{\theta_g}{L_G }
        \]
        \ENDFOR
\ENDFOR
\end{algorithmic}
\end{algorithm}

\subsubsection{Binary Cross Entropy Loss.}
Binary Cross Entropy (BCE) is one of the most widely used loss functions for training the generator in GANs because it effectively quantifies the difference between two data distributions. BCE assesses how successfully the generator has deceived the discriminator into classifying the generated outputs as real rather than synthetic. When employing BCE loss for the generator, the labels for its outputs are flipped to 1, indicating ``real.'' Thus, the generator aims to minimize the following loss:

\begin{equation}
    L_G = -\frac{1}{m} \sum_{i=1}^{m} \log(D(G(z_i))) 
\end{equation}


where \( m \) denotes the batch size.

\subsubsection{Feature Matching Loss.}
During the discriminator’s forward pass with real and synthetic malware samples, intermediate features from a middle hidden layer are extracted. A distance metric then calculates the difference between the average features of real and synthetic malware samples. The generator's loss function is defined as:

\begin{equation}
    L_G =  \frac{1}{m} \sum_{i=1}^{m} \| \mathbb{E}_{x \sim p_{\text{data}}}[D^{(f)}(x)] - \mathbb{E}_{z \sim p_z}[D^{(f)}(G(z))] \| 
\end{equation}

where \( D^{(f)}(\cdot) \) denotes the output of a middle layer of the discriminator, \( x \) represents real malware samples, \( G(z) \) represents synthetic malware samples, and \( m \) denotes the batch size.

This loss function, adopted by previous research~\cite{salimans2016improved}, provides a compelling alternative to BCE loss by refocusing training objectives from the final discriminator output to the richer intermediate feature layers, fostering a more nuanced learning process for the generator.



$$$$

\section{Experimental Details}
\label{sec:eval}
In this section, we detail the datasets and features used in the evaluations, introduce baseline models and those from prior literature~\cite{gr,bir} for comparison with MalCL, and discuss the experimental settings to demonstrate the effectiveness of MalCL as a malware family detection methodology.


\subsection{Datasets}
\label{subsec:data}
We use two large-scale malware datasets for our experiments: EMBER~\cite{ember}, which includes Windows PE malware samples, and AZ-Class, a dataset we collected from the AndroZoo repository~\cite{AndroZoo} containing Android malware samples. Consistent with prior work~\cite{droidevolver}, we selected samples with a VirusTotal detection count $>=$ four.

\subsubsection{EMBER.} 

We use the 2018 EMBER dataset, known for its challenging classification tasks, focusing on a subset of 337,035 malicious Windows PE files labeled by the top 100 malware families, each with over 400 samples. Features include file size, PE and COFF header details, DLL characteristics, imported and exported functions, and properties like size and entropy, all computed using the feature hashing trick.


\subsubsection{AndroZoo.} 



The AZ-Class dataset contains 285,582 samples from 100 Android malware families, each with at least 200 samples. We extracted Drebin features~\cite{arp2014drebin} from the apps, covering eight categories like hardware access, permissions, API calls, and network addresses. Initially, the training set had 1,067,550 features, which were aligned in the test set. We used \texttt{scikit-learn’s VarianceThreshold} to reduce low-variance features, resulting in a final feature dimension of 2,439.


\subsection{Baseline and Prior Work}



This work uses two baselines: \textit{None} and \textit{Joint}. In the \textit{None} baseline, the classifier is trained only on the new data for each task, simulating catastrophic forgetting (CF) and serving as the informal lower bound. The \textit{Joint} baseline combines data from all previous tasks into a single dataset for training, representing an ideal scenario that preserves all prior information but is impractical for large-scale malware datasets, serving as the informal upper bound.



The baseline results are empirically grounded. A standard scaler is applied in all scenarios, with variations in its usage: it is refitted for each new task in the None and Joint baselines but incrementally updated in CL scenarios. This highlights a key distinction: the None baseline trains exclusively on new data, the Joint baseline integrates all data, and CL scenarios adaptively learn evolving data distributions as new classes are introduced.


We also compare the performance of MalCL with two prior GR-based CL techniques: GR~\cite{gr} and BI-R~\cite{bir}, both well-studied in CV  domains. GR uses a GAN to GR samples, while BI-R, which employs a Variational Autoencoder (VAE)~\cite{kingma2013auto}, enhances GR by adding Conditional Replay (CR), Gating based on Internal Context (Gating), and Internal Replay (IR).


\begin{table}[!t]
\small
\begin{center}
\caption{Comparisons to Baseline and Prior Replay Models Using  Ember Dataset. We report the mean accuracy scores (Mean) and minimum (Min) computed from every 11 tasks.}
\label{tab:comp}
\vspace{-0.3cm}
\setlength{\tabcolsep}{6pt}
\renewcommand{\arraystretch}{1.2}
\begin{tabular}{|l|ll|ll|}
\hline
\multicolumn{1}{|c|}{\multirow{2}{*}{\textbf{Approach}}} & \multicolumn{2}{c|}{\multirow{2}{*}{\textbf{Method}}} & \multicolumn{2}{c|}{\textbf{EMBER}} \\ \cline{4-5} 
\multicolumn{1}{|c|}{} & \multicolumn{2}{c|}{} & \multicolumn{1}{c}{\textit{\textbf{Mean}}} & \multicolumn{1}{c|}{\textit{\textbf{Min}}} \\ \hline
\multirow{2}{*}{Baselines} & \multicolumn{2}{l|}{None} & 27.5 & 0.6 \\
 & \multicolumn{2}{l|}{Joint} & 88.7 & 74.5 \\ \hline

\multirow{2}{*}{Prior Works} & \multicolumn{2}{l|}{GR} & 26.8 & 09.5 \\
 & \multicolumn{2}{l|}{BI-R} & 27.0 & 09.2 \\ \hline

\multirow{6}{*}{MalCL} & \multicolumn{1}{l|}{\multirow{2}{*}{\begin{tabular}[c]{@{}l@{}}L2 to Labels\end{tabular}}} & FML & 50.2 & 17.8 \\
 & \multicolumn{1}{l|}{} & BCE & 47.8 & 19.4 \\ \cline{2-5}

 & \multicolumn{1}{l|}{\multirow{2}{*}{\begin{tabular}[c]{@{}l@{}}L1 to CMean \\ Logits\end{tabular}}} & FML & \textbf{54.5} & 21.8 \\
 & \multicolumn{1}{l|}{} & BCE & 52.1 & 26.2 \\ \cline{2-5} 
 & \multicolumn{1}{l|}{\multirow{2}{*}{\begin{tabular}[c]{@{}l@{}}L1 to BMean \\ Logits\end{tabular}}} & FML & 50.3 & 32.0 \\
 & \multicolumn{1}{l|}{} & BCE & 47.6 & 24.2 \\ \hline
\end{tabular}
\end{center}
\vspace{-0.3cm}
\end{table}

\subsection{Task Set Construction and Metrics}
We constructed the training set for the first task by randomly selecting 50 classes, adding 5 randomly chosen classes for each subsequent task, resulting in 11 task datasets. Each dataset simulates incremental additions of malware classes to the training set. For the testing set, we used all available classes observed up to each task -- for instance, 50 classes for the first task, 55 for the second, and so forth. To evaluate the impact of random selection, we conducted five experiments, reporting the minimum and mean accuracies achieved. Figures~\ref{fig:overall} and~\ref{fig:dataset} illustrate these accuracies; each line represents mean accuracies, while the shaded areas reflect the range between the maximum and minimum accuracies at each step.

\section{Results}
In this section, we discuss the performance of MalCL by comparing it to baseline models and those from prior literature~\cite{gr,bir}, exploring various generator loss functions, replay sample selection methods, and task set generation scenarios.

\begin{figure}[!t]
    \centering
    \begin{subfigure}[b]{0.475\linewidth}
        \includegraphics[width=\linewidth]{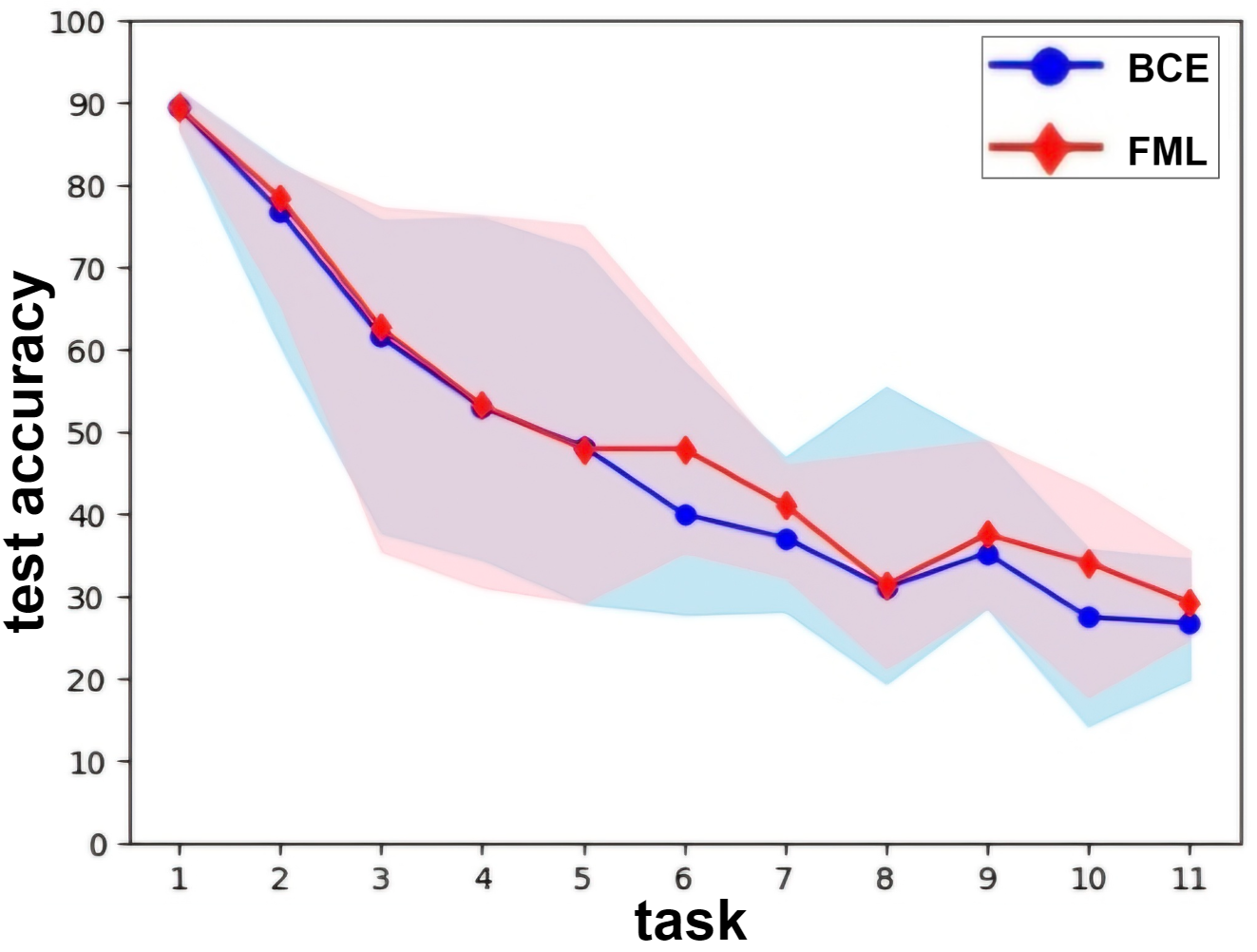}
        \caption{L2 to Labels: 48\% vs. 50\%.}
        \label{fig:logitClass}
    \end{subfigure}
    \begin{subfigure}[b]{0.475\linewidth}
        \includegraphics[width=\linewidth]{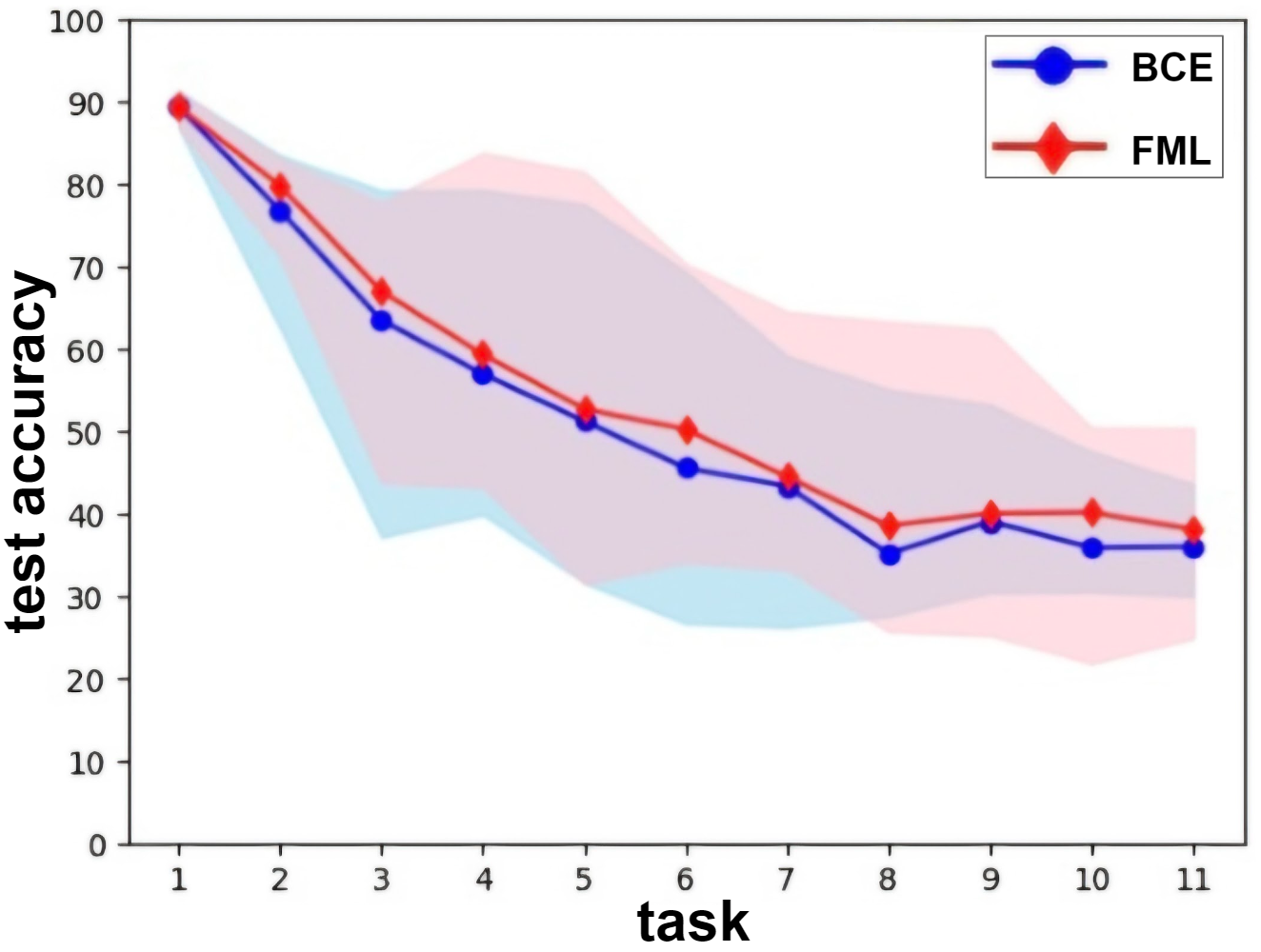}
        \caption{L1 to CMean: 51\% vs. 55\%.}
        \label{fig:softmax}
    \end{subfigure}

    \caption{Impact of three replay sample selections on MalCL performance when using the EMBER Dataset. We further report the mean accuracies for each task when comparing BCE with FML.}
    \label{fig:overall}
    \vskip -0.3cm
\end{figure}

\subsection{Comparisons to Baseline and Prior Work}



MalCL demonstrated superior performance compared to the None baseline, GR, and BI-R. In the optimal setting, where FML with L1 distance to Per-Class Mean Logit Vectors (L1 to CMean Logits) was employed, MalCL improved upon the None baseline by 27\%. This substantial performance gain indicates that MalCL is significantly more robust in mitigating CF, a critical issue observed in the None baseline. Furthermore, MalCL outperformed existing CL approaches with replay, such as GR and BI-R, by 28\%, highlighting that while these methods are effective in image classification tasks, they are less effective in classifying malware families when new malware samples are continually introduced. Our observations also revealed that CL techniques optimized for image classification, such as GR and BI-R, performed similarly to the None baseline, indicating their limited effectiveness in addressing CF in malware classification. This underscores the necessity of designing and optimizing CL methodologies specifically targeted at malware family classification.

\begin{figure}[!t]
    \centering
    \begin{subfigure}[b]{0.475\linewidth}
        \includegraphics[width=\linewidth]{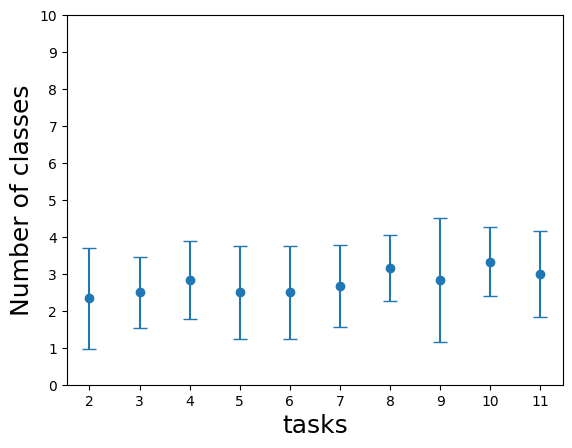}
        \caption{BCE}
        \label{fig:class_cnt_BCE}
    \end{subfigure}
    \hfill
    \begin{subfigure}[b]{0.475\linewidth}
        \includegraphics[width=\linewidth]{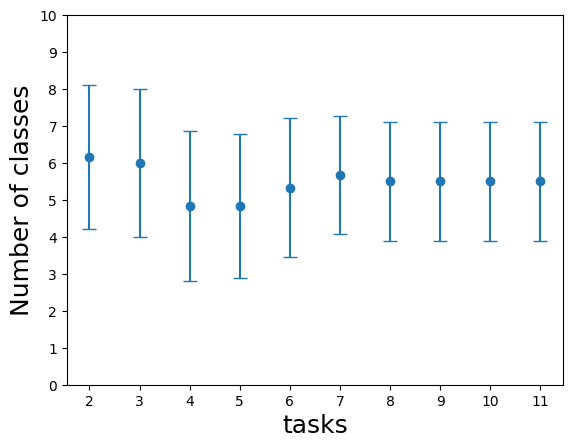}
        \caption{FML}
        \label{fig:class_cnt_FML}
    \end{subfigure}

    \caption{The number of replay sample classes per task using L1 to BMean Logits. We omit the first task, as no replay sample is utilized to train the classifier.}
    \label{fig:overall}
    \vskip -0.3cm
\end{figure}

However, we also note that the selection scheme based on L1 distance to Per-Batch Mean Logit Vectors (L1 to BMean Logits) demonstrated poor performance, producing synthetic samples for only a few classes and thus restricting the replay ability of MalCL (Refer to Figure~\ref{fig:overall}). This scheme does not explicitly enforce sample selection per class, leading to the natural inability of the generator to diversify its production to cover all malware families. We plan to work on improving the global selection scheme in the future.

In addition, although there remains a substantial gap between the performance of MalCL and the Joint baseline, there is potential for further improvement, particularly in more accurately emulating original samples and generating even more robust synthetic malware samples. Enhancing these aspects could bring performance closer to that of the Joint baseline, which we identify as a promising direction for future research.

\subsection{Generator Loss and Replay Sample Selection}
\label{subsec:comp_gloss}




We compare the performance of two selection schemes: L2 to Labels and L1 to CMean Logits, using both BCE and FML. As depicted in Figure~\ref{fig:logitClass}, FML slightly outperforms BCE across all tasks for both schemes.

For both loss functions, MalCL demonstrates increased effectiveness with the L1 to CMean Logits selection scheme. This improvement indicates that synthetic malware samples, which closely resemble the mean vectors derived from the richer features of the classifier's hidden layer, effectively facilitate the replay of training data from prior tasks. This sampling scheme aligns well with the generator model's objective of accurately mimicking original malware samples, as captured by mean logit vectors. Consequently, these replay samples are likely to exhibit lower classification errors when identified as the target class. Overall, FML with L1 to CMean Logits achieves the highest accuracy across all tasks in classifying malware families.

\subsection{Task Set Generation Methods}
The method of constructing task sets is crucial for the CL capabilities of MalCL. While we have employed random selections in all previously described experiments, we also explored a scenario where larger classes containing more malware samples were assigned to the initial tasks, and smaller classes with fewer samples to subsequent tasks. In this configuration, we fixed the first task to include 50 ``giant'' classes, each having an average sample count of 5,397, and applied random selection to the remaining classes, which have a mean sample count of 670, to construct the sets for later tasks. Surprisingly, MalCL, with the EMBER dataset, replay sample selection scheme of L1 to CMean Logits and BCE loss, achieved a mean accuracy of 74\%, significantly surpassing the previous best result of 55\% and closely approaching the performance of the Joint baseline (refer to Table~\ref{tab:comp}). This outcome underscores that assigning large classes with a greater number of samples to the first task can effectively mitigate the impact of CF on earlier task data in the CL process.

\begin{figure}[!t]
   \centering
    \includegraphics[width=0.6\linewidth]{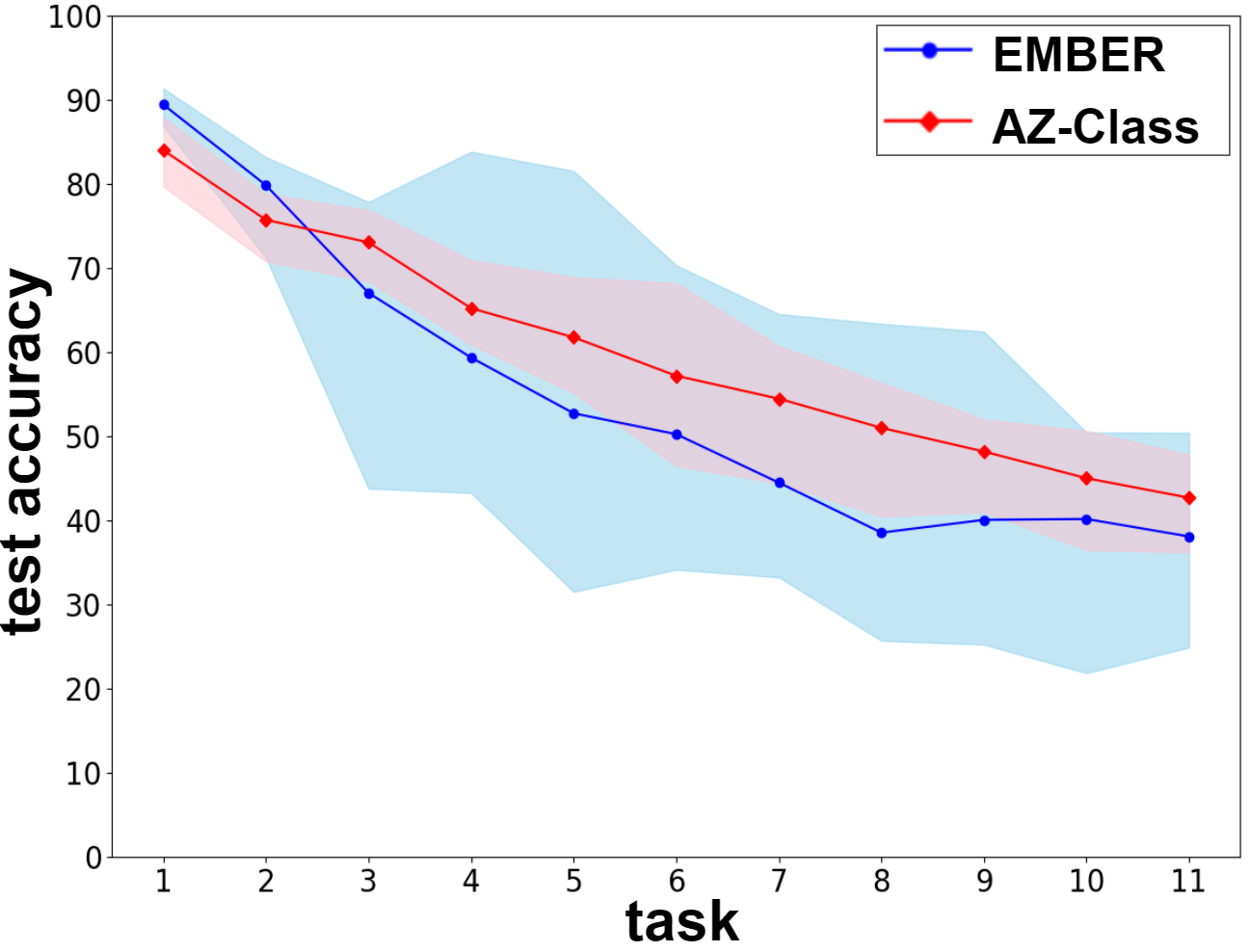}
    \caption{MalCL Performance on the EMBER and AZ-Class datasets using FML and L1-norm to Mean Logits.}
    \label{fig:dataset}
    \vskip -0.3cm
\end{figure}


\subsection{MalCL on AZ-Class}

We further evaluated MalCL using the AndroZoo dataset (AZ-Class) and observed, as illustrated in Figure~\ref{fig:dataset}, that MalCL performed better compared to when using the EMBER dataset.


Additionally, the gap between the minimum and maximum accuracy ranges is narrower. This improvement can be attributed to the EMBER dataset's more imbalanced set, characterized by a highly variable sample count per class. Such imbalance weakened MalCL's ability to retain learning from previous tasks and effectively learn from new observations, which added confusion to the malware family classification.

We summarize key findings through our experiments as follows.

\begin{itemize}
    \item MalCL improved upon the None baseline by 27\% and surpassed GR and BI-R by 28\%, underscoring its effectiveness in malware classification compared to models optimized for other domains like image classification.
    \item The effectiveness of MalCL’s performance is closely linked to the alignment of replay samples with the original data’s feature space as defined by the classification model, especially when employing the L1 distance to Per-Class Mean Logits selection scheme.
    \item Assigning larger classes with more samples to earlier tasks significantly improved MalCL’s accuracy (up to 74\%), highlighting the importance of strategic task set construction in CL scenarios.
    \item The evaluation on the EMBER and AZ-Class datasets revealed that dataset imbalance can weaken MalCL’s learning retention and accuracy, suggesting the need for strategies to handle class imbalances effectively in CL models.
\end{itemize}


\section{Conclusion}
MalCL achieves state-of-the-art performance in mitigating CF in malware classification across Windows and Android platforms, underscoring the efficacy of our GR-based CL techniques. Our research robustly confirms that these methods can be effectively adapted to the malware domain. Looking ahead, improving the quality of synthetic malware generation remains critical. We aim to develop more advanced generative models and investigate hybrid training approaches that merge the advantages of GR and joint training. Future research will expand to address a wider variety of malware types and integrate more sophisticated features that reflect the dynamic nature of malware threats. Additionally, we will explore adaptive mechanisms to proactively anticipate and counter shifts in malware tactics, further bolstering MalCL's robustness.
\section*{Acknowledgments}
We would like to thank our anonymous reviewers for helpful comments. This work was supported by Institute of Information \& communications Technology Planning \& Evaluation (IITP) grant funded by the Korea government (MSIT) (No. RS-2022-00155966, Artificial Intelligence Convergence Innovation Human Resources Development (Ewha Womans University)), and by  the National Research Foundation of Korea (NRF) grant funded by the Korea government (MSIT) (No. RS-2023-00222385).


\bibliography{main}

\end{document}